\documentclass[]{spie}  
\usepackage[]{graphicx}
\usepackage{multirow}

\title{Generation of spatially pure photon pairs in a multimode nonlinear
waveguide using intermodal dispersion}

\author{Micha{\l} Karpi{\'n}ski, Czes{\l}aw Radzewicz, and Konrad Banaszek
\skiplinehalf
Faculty of Physics, University of Warsaw, ul.\ Ho{\.z}a 69, 00-681 Warszawa, Poland
}

\authorinfo{Further author information -- send correspondence to:\\M.K.: E-mail: michal.karpinski@fuw.edu.pl, Telephone: +48 22 55 32330
 }

\begin{document}
  \maketitle

\begin{abstract}
We present experimental realization of type-II spontaneous parametric down-conversion in a periodically poled potassium titanyl phosphate (KTiOPO$_4$) nonlinear waveguide. We demonstrate that by careful exploitation of intermodal dispersion in the waveguide it is feasible to produce photon pairs in well defined transverse modes without any additional spatial filtering at the output. Spatial characteristics is verified by measurements of the $M^2$ beam quality factors. We also prepared a postselected polarization-entangled two-photon state shown to violate Bell's inequality. Similar techniques based on intermodal dispersion can be used to generate spatial entanglement and hyperentanglement.
\end{abstract}



\section{INTRODUCTION}
\label{sec:intro}

Generating nonclassical optical radiation attracts currently a lot of interest owing to its prospective use in quantum-enhanced technologies, such as quantum cryptography \cite{QCrypto1, QCrypto2}, precision measurements \cite{Metrology}, and other emerging applications \cite{OBriFuruNPH09}. A common approach to generate nonclassical radiation is to realize spontaneous parametric down-conversion (SPDC) in a $\chi^{(2)}$ nonlinear optical medium,\cite{ProgOpt} which is easy to set up and can be operated at room temperatures. However, standard bulk media suffer from low conversion efficiencies and typically generate light that exhibits a complicated spatio-temporal structure. In contrast, many protocols require single photons and photon pairs prepared in well defined spatial and spectral modes in order to facilitate high visibility multiphoton interference\cite{URenBanaQIC03} as well as efficient coupling into optical fibers and integrated optical circuits.\cite{PoliMattIEEE09} A standard solution to this problem is to implement spatial and spectral filtering to ensure suitable coherent structure of optical fields. However, such an approach typically introduces substantial losses\cite{KoleWasiPRA09} which deteriorate overall performance and destroy photon number correlations that are important in certain protocols.

A promising alternative solution to the above problems is to resort to nonlinear fibers and waveguides.\cite{FiberSources1, FiberSources2, FiberSources3, FiberSources4, FiberSources5, FiberSources6, FiberSources7, FiberSources8} In an optical guided structure, spatial confinement of optical fields to its transverse cross-section enhances the strength of a nonlinear interaction. At the same time, the confinement changes dramatically the spatial characteristics of the produced light. This offers a possibility to produce photons in required transverse modes right at the source. In addition, these modes can be directly matched to single-mode fibers,\cite{PPKTP3} 
 ensuring high coupling efficiency.

A popular material for making nonlinear waveguides is potassium titanyl phosphate (KTP),\cite{PPKTP1, PPKTP2, PPKTP3, PPKTP4} which with the technique of periodic poling allows one to cover the 400~nm $\rightarrow$ 800~nm + 800~nm spectral region through quasi-phase matching. However, standard manufacturing process produces periodically poled (PP) KTP waveguides that are multimode at both 400~nm pump and 800~nm down-converted wavelengths. This potentially compromises the ability to generate photon pairs in well defined spatial modes. Fortunately, this turns out not to be the case: intermodal dispersion, which in standard telecom applications is seen mainly as a nuisance, provides a useful tool to control nonlinear process in optical waveguides.\cite{AndersonConf71, AndersonAPL71} In the case of the parametric down-conversion process, this was demonstrated in our recent experiment\cite{KarpRadzOL12} in which spatially pure photon pairs were generated in a multimode waveguide. Achieving these results relied essentially on our previous experimental effort to characterize the nonlinear three-wave mixing process in the wave guiding structure\cite{KarpRadzAPL09}.

In this contribution we review the principle of exploiting intermodal dispersion for controlling spatial modes in waveguided parametric down-conversion and present its experimental verification. We also report on utilization of the constructed source to generate polarization entangled photon pairs in the Shih-Alley configuration\cite{ShihAllePRL88}, with observed polarization visibilities reaching values that violate Bell's inequalities. These results demonstrate the feasibility of engineering the modal structure of nonclassical radiation in nonlinear waveguides, opening up a feasible route towards their use in practical applications.

\section{METHOD}
\label{sec:disp}

\begin{figure}
\centering
\includegraphics[width=12cm]{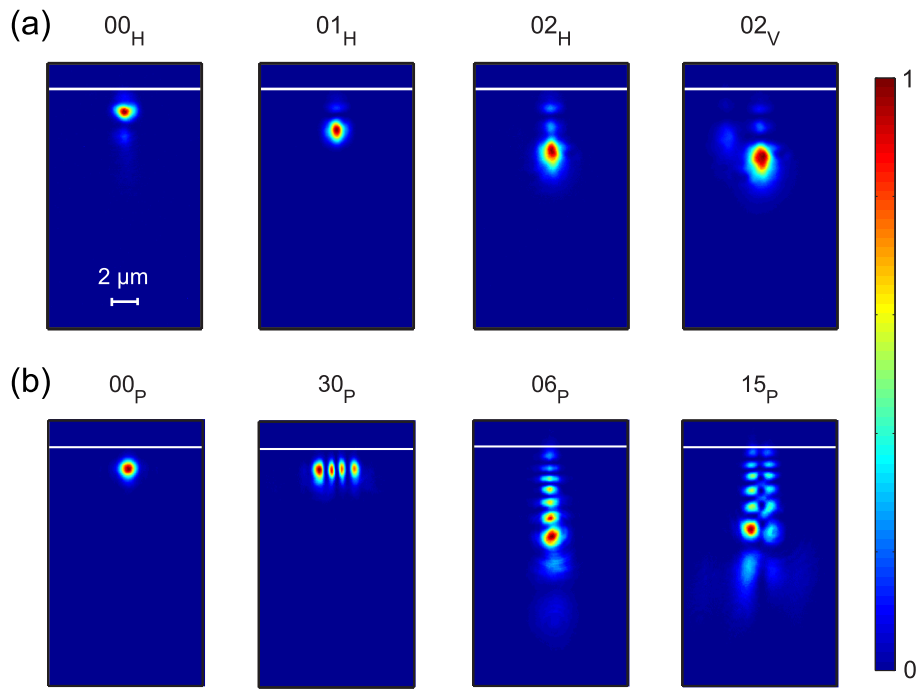}
\caption{Intensity distributions imaged from the output face of the waveguide corresponding to transverse modes excited by macroscopic $800$~nm (a) and $400$~nm (b) beams. For the labelling of the transverse modes, see the main text. In the upper rightmost panel, traces of a ghost reflection from imaging optics can be seen.}
\label{Fig:modes}
\end{figure}

Let us begin by describing the principle of controlling the spatial characteristics of down-converted radiation using intermodal dispersion in a more detailed manner than presented in our earlier communications\cite{KarpRadzAPL09,KarpRadzOL12}.

For concreteness, we will focus on a type-II spontaneous parametric down-conversion process in which a pump photon, labeled with an index $P$, decays into a pair of orthogonally polarized photons denoted as $H$ and $V$. In a multimode waveguide, each of the fields $P$, $H$, and $V$ can be prepared in a number of different spatial modes. This is illustrated with Fig.~\ref{Fig:modes} which presents images, obtained using a microscope objective and a CCD camera, of transverse intensity distributions for different spatial modes that have been excited in the waveguide using macroscopic coherent beams. We will label individual modes with a pair of integers $ij$ specifying the number of nodes in two orthogonal directions and a subscript index corresponding to the field $P$, $H$, or $V$ which defines the polarization and the frequency region.

The spectral properties of SPDC photons are governed by energy and momentum conservation. Energy conservation means that the frequencies $\omega_H$ and $\omega_V$ of the generated photons must sum up to the frequency of the pump photon. Momentum conservation is known in nonlinear optics as the phase matching condition. In bulk nonlinear media the phase matching condition is defined by the wave vectors of the three interacting fields. For the collinear configuration in a periodically poled medium it takes the form
\begin{equation}
k_P(\omega_H+\omega_V) = k_H(\omega_H) + k_V(\omega_V) + \frac{2\pi n}{\Lambda},
\end{equation}
where $k_P$, $k_H$, and $k_V$ are wave vectors for the respective fields, $\Lambda$ is the poling period, and $n$ is the quasi-phase matching order. In a waveguide wave vectors acquire dependence on the transverse mode propagating through the structure. This effect is known as intermodal dispersion. The wave vector for a specific transverse mode $k_\mu^{ij}$ ($\mu=P, H, V$) can be written as a sum of that for the bulk medium $k_\mu (\omega_\mu)$ and a mode dependent contribution $\Delta k^{ij}$, which for sufficiently narrow spectral ranges can be assumed to be frequency-independent:
\begin{equation}
k_\mu^{ij} (\omega_\mu) = k_\mu (\omega_\mu) + \Delta k^{ij}, \qquad  \mu=P, H, V ,
\end{equation}
hence $\Delta k^{ij}$ can be referred to as the geometric contribution. As a result,
the phase matching condition of a down-conversion process for a specific triplet of modes $lm_P \rightarrow ij_H + i' j' _V$
\begin{equation}
k_P^{lm}(\omega_H+\omega_V) = k_H^{ij}(\omega_H) + k_V^{i'j'}(\omega_V) + \frac{2\pi n}{\Lambda},
\end{equation}
is typically satisfied for different sets of frequencies.

\begin{figure}
\centering
\includegraphics[width=17cm]{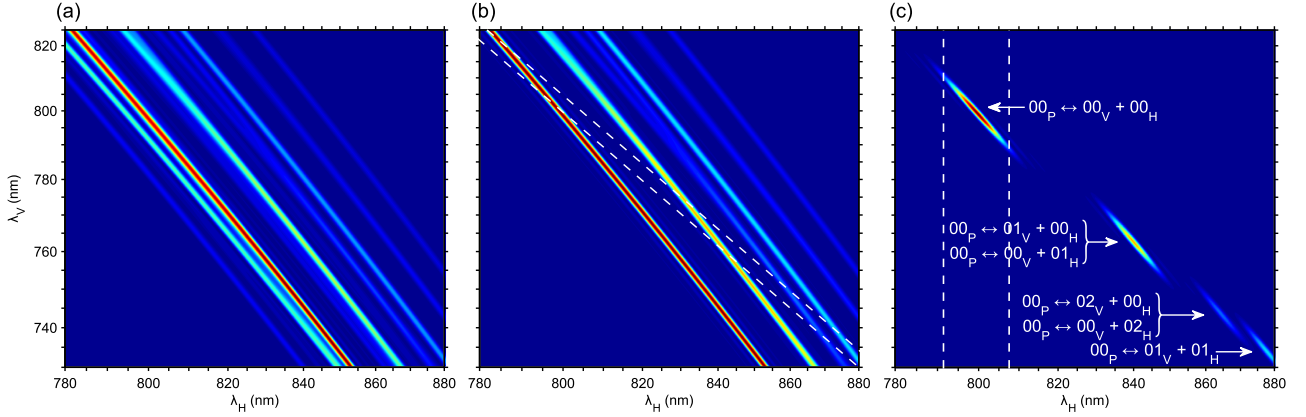}
\caption{The squared absolute value of phase matching bands for a $L=1$~mm long the PPKTP waveguide under investigation assuming (a) multimode pump and (b) the pump field mode fixed to $00_P$. The phase matching functions have been calculated using experimental data measured for the sum-frequency generation process at the degenerate wavelength. Individual bands correspond to different pairs of down-conversion modes for photons $H$ and $V$. In (b), the pair of dashed lines bounds the energy conservation region defined by the pump spectrum. Combination of the phase matching constraints and the pump spectrum yields the joint spectrum of down-coverted photon pairs shown in (c). Each spectral island is identified with triplets of transverse modes contributing to the down-conversion process in a given spectral region. The pair of pure fundamental spatial modes $00_H + 00_V$ can be selected by a coarse filtering of one of the photons, as shown schematically with vertical dashed lines. For convenience, the axes, scaled in frequencies, are labelled with vacuum wavelengths of the down-converted photons.}
\label{Fig:maps}
\end{figure}

In a nonlinear medium of finite length the phase matching condition does not have to be satisfied exactly. The probability amplitude of converting a pump photon into a pair of photons with frequencies $\omega_H$ and $\omega_V$ is given by the phase matching function proportional to
\begin{equation}
\Phi_{lm_P \rightarrow ij_H + i' j' _V}(\omega_H,\omega_V) \propto {\rm sinc} \left[ \frac{L}{2} \left( k_P^{lm}(\omega_H+\omega_V) - k_H^{ij}(\omega_H) - k_V^{i'j'}(\omega_V) - \frac{2\pi n}{\Lambda} \right) \right]
\end{equation}
where $L$ is the medium length. The proportionality factor includes the efficiency of the process which depends on the spatial overlap of the interacting transverse modes. The collection of phase matching functions for a PPKTP waveguide parameterized with wavelengths of the down-converted photons is shown in Fig.~\ref{Fig:maps}(a). It is clearly seen that intermodal dispersion separates phase matching functions for different triplets of $P$, $H$, and $V$ modes into distinct bands. This is a crucial feature that will allows us to control the transverse modes of the down-converted photons.

When realizing the  down-conversion process, only bands involving the pump mode that has been excited in the waveguide are relevant. If our goal is to produce pairs of photons in the fundamental spatial modes $00_H$ and $00_V$,
the efficiency of the down-conversion process is highest for the pump beam  prepared in the $00_P$ mode thanks to maximized overlap of transverse modes.\cite{FiorMunroOE07} 
This conclusion is directly confirmed by measurements of the sum-frequency process and their comparison with numerical modeling,\cite{KarpRadzAPL09} summarized in Tab.~\ref{tab:waski_lista}. We will call the phase matching band corresponding to the $00_P \rightarrow 00_H + 00_V$ process the {\em fundamental band}. The intermodal dispersion makes wave vectors corresponding to  higher spatial $H$ and $V$ modes differ from the $00_H$ and $00_V$ by the same sign. This makes the fundamental band separated from bands $00_P \rightarrow ij_H+ i'j'_V$ that involve other than fundamental combinations of down-converted modes, as shown in Fig.~\ref{Fig:maps}(b)  (see also Sec.~\ref{sec:exp_setup} and Tab.~\ref{tab:waski_lista}). The next step is to consider the spectrum of the pump, which defines a strip within in the wavelength plane of the down-converted photons within which the energy of the SPDC process is conserved. Because the photons are generated via a type-II process in a birefringent medium, the slope of the phase matching bands in Fig.~\ref{Fig:maps}(a,b) is typically other than $45^\circ$, while the strip defined by the pump spectrum is symmetric with respect to swapping the wavelengths of the down-converted photons. As a result, the pump strip crosses several bands, creating a number of islands shown in Fig.~\ref{Fig:maps}(c). For the PPKTP waveguide considered here, the islands are sufficiently separated to isolate the $00_P \rightarrow 00_H + 00_V$ by coarse spectral filtering of one of the produced photons. This leaves us with a single pair of fundamental spatial modes, thus enabling generation of spatially pure photon pairs by down-conversion in a multimode waveguide.

\begin{table}
\centering
\vspace{5mm}
\begin{tabular}{ccccc}
\hline
$\lambda$~(nm) & $ij_V$ & $ij_H$ & $ij_P$ & Relative efficiency$(\%)$\\
\hline
793.10 & 10 & 10 & 20 & 10\\ 
795,00 & 00 & 01 & 01 & 3 \\ 
796.60 & 01 & 01 & 02 & 3 \\ 
796.70 & 10 & 00 & 10 & 54 \\ 
797.10 & 02 & 01 & 02 & 3 \\ 
798.50 & 02 & 00 & 01 & 15 \\ 
798.70 & 00 & 10 & 10 & 23 \\ 
\textbf{799.40} & \textbf{00} & \textbf{00} & \textbf{00} & \textbf{100} \\ 
800.10 & 00 & 02 & 01 & 13 \\ 
802.20 & 02 & 02 & 02 & 3 \\ 
804.40 & 10 & 01 & 10 & 3 \\ 
806.00 & 01 & 10 & 10 & 13 \\ 
\textbf{806.80} & \textbf{01} & \textbf{00} & \textbf{00} & \textbf{40} \\ 
\textbf{807.60} & \textbf{00} & \textbf{01} & \textbf{00} & \textbf{25} \\ 
\textbf{811.50} & \textbf{02} & \textbf{00} & \textbf{00} & \textbf{12} \\ 
\textbf{813.30} & \textbf{00} & \textbf{02} & \textbf{00} & \textbf{8} \\ 
\textbf{815.00} & \textbf{01} & \textbf{01} & \textbf{00} & \textbf{20} \\ 
\hline
\\
\end{tabular}
\caption{A list of experimentally identified frequency-degenerate SFG processes for the PPKTP waveguide. Processes involving the fundamental $00_P$ pump mode are typeset in boldface. Experimentally determined relative efficiencies are specified with respect to the fundamental process.}
\label{tab:waski_lista}
\end{table}

\section{EXPERIMENTAL SETUP}
\label{sec:exp_setup}

\begin{figure}[t!] 
\centering
\includegraphics[width=16cm]{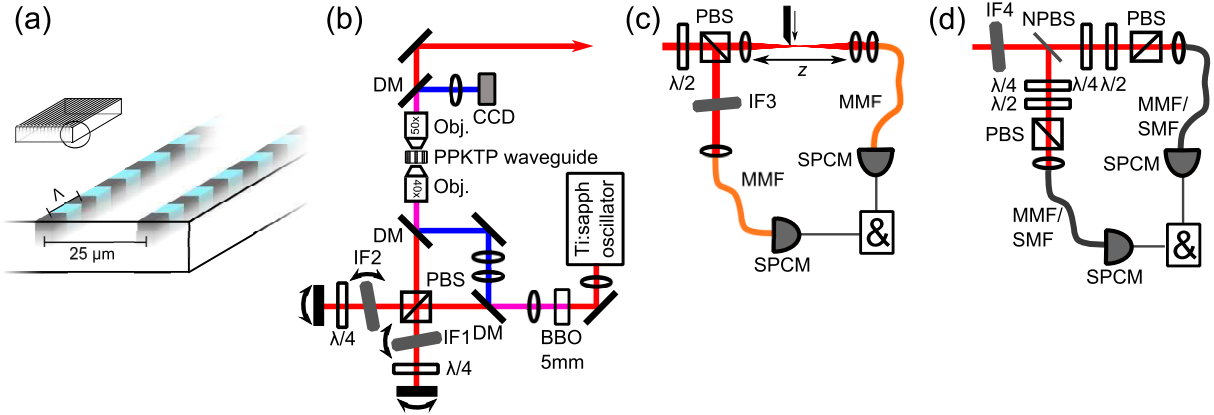}
\caption{(a) The waveguide sample used in the experiment. A diagram of the photon pair source (b) and the setups to measure the beam quality factors (c) and polarization correlations (d). DM, dichroic mirrors; $\lambda/4$, quarter-wave plates; PBS, polarizing beam splitters. For other symbols see the main text.
}
\label{Fig:setup}
\end{figure}

The nonlinear waveguide used in our experiment was a 1~mm long PPKTP structure (AdvR Inc.) shown in Fig.~\ref{Fig:setup}(a) which contained a number of ion-exchanged waveguides with a poling period of $7.5~\mu$m, suitable for type-II down-conversion with radiation generated in the 800 nm wavelength region.  In the transverse plane, the waveguides were 2~$\mu$m wide in the direction parallel to the crystal surface and had an approximately exponentially decaying refractive index profile in the perpendicular direction, characterized by the effective depth of ca.\ $5~\mu$m.  The mode labels $ij$ refer respectively to the number of nodes in these two directions, and the polarization $H$ is defined as parallel to the crystal surface.

The experimental setup to characterize the waveguide and to generate photon pairs is shown schematically in Fig.~\ref{Fig:setup}(b).
Because the phase matching function plays the essential role in our scheme for generating spatially pure photon pairs, we characterized it directly before assembling the down-conversion source. This was done
using the mode-resolved sum frequency generation (SFG) spectroscopy technique described and implemented in our earlier work\cite{KarpRadzAPL09}. Two 800~nm region infrared (femtosecond) pulsed beams with $H$ and $V$ polarizations were tuned by rotating $0.6$~nm FWHM interference filters IF1 and IF2 and launched into the waveguide in independently controlled spatial modes. The sum frequency intensity and its spatial distribution were recorded on the CCD camera. Using this setup we could obtain a spectral location of phase matching bands for degenerate frequencies of the $H$ and $V$ fields. The most efficient bands observed in the $795$ -- $815$~nm region are listed in Tab.~\ref{tab:waski_lista}.
In particular, we found the center of the fundamental band  at $799.4$~nm with $1.0$~nm FWHM bandwidth. In agreement with theoretical predictions, the closest band involving $00_P$ mode and higher $H$ and/or $V$ modes was found to be separated by more than $6.0$~nm. Bands corresponding to $10_P \leftrightarrow 10_V + 00_H$, $10_P \leftrightarrow 00_V + 10_H$, $01_P \leftrightarrow 02_V + 00_H$, and $01_P \leftrightarrow 00_V + 02_H$ processes were found in the close vicinity (less than $3$ nm separation) of the fundamental band. This meant that selecting the fundamental band relied critically on coupling the pump beam into $00_P$ mode and minimizing the excitation of $10_P$ and $01_P$ modes. After characterizing the waveguide, we tuned the wavelengths of the red beams to maximize the sum frequency signal from the fundamental band. These auxiliary beams were later used to align and calibrate the detection setup.

To produce photon pairs, a 399.9~nm pump beam with 1.0~nm FWHM bandwidth, obtained by frequency doubling 150~fs pulses from a Ti:sapphire oscillator (Coherent Chameleon Ultra) in a 5~mm long beta barium borate (BBO) crystal, was coupled into the waveguide. The spatial mode of the beam was matched to the $00_P$ waveguide mode by a combination of lenses, a zoomable beam expander to fine-tune the size of the beam at focus, and a microscope objective. The light exiting the waveguide was collected by a $0.8$ numerical aperture (NA) birefringence-free microscope objective. The pump beam was filtered out by dichroic mirrors and directed to a CCD camera imaging the output face of the waveguide in order to continuously monitor the spatial mode of the pump beam. We performed a careful comparison of the excited pump beam mode with the reference $00_P$ intensity distribution obtained directly from sum-frequency generation. The intensity distributions of the excited mode and the SFG signal were used to calculate the integral of the product of the corresponding normalized mode functions. Assuming constant phase, the obtained value of the overlap was equal to $0.966 \pm 0.002$.

\section{BEAM QUALITY MEASUREMENTS}

The spatial properties of the photons produced in the waveguide were characterized using the standard approach to determine the $M^2$ beam quality parameter via free-space diffraction\cite{QuadraticPropagation1, QuadraticPropagation2}. The challenge was to realize the measurement at the single-photon level and to condition data upon the detection of the heralded conjugate photons.

In the experiments reported in this section, the waveguide was typically pumped with  $50~\mu$W of power. This figure corresponds to the power within the waveguide, with coupling efficiency exceeding $50\%$.
Details of the experimental apparatus to characterize the spatial coherence of generated photons are shown in Fig.~\ref{Fig:setup}(c).
First, $H$ and $V$ photons were separated on a polarizing beam splitter. A half-wave plate placed before the polarizer was used to switch the $H$ and $V$ beams between the outputs. The transmitted photons, whose spatial characteristics was to be measured, were focused with a $150$~mm focal length lens to a waist of approximately $100~\mu$m, corresponding to the Rayleigh range $z_R \approx 40$~mm. The transverse distribution was probed with the edge of a knife mounted on a motorized translation stage with a $0.2~\mu$m positioning resolution. The edge could be placed in different locations $z$ in the beam propagation direction by moving it along an optical rail, and swept through the beam in the transverse plane in either horizontal or vertical direction. To comply with the ISO standard of $M^2$ value measurement\cite{ISOM2}, we located the knife edge at a minimum of 5 points within the Rayleigh range, and at least 5 points outside $2z_R$. The light passing the edge was recollimated by another 150~mm lens, and focused with an $8$~mm focal length aspheric lens (AL) into a $100~\mu$m core diameter multimode fiber (MMF) that delivered the signal to a single photon counting module  (SPCM-AQRH, Perkin Elmer). We verified that higher-order waveguide modes were also coupled into the MMF by exciting them with the auxiliary red beam and observing the intensity distribution at the focus of the AL with a camera and simultaneously confirming their good coupling efficiency into multimode fibers.

\begin{figure}
\centering
\includegraphics[width=16cm]{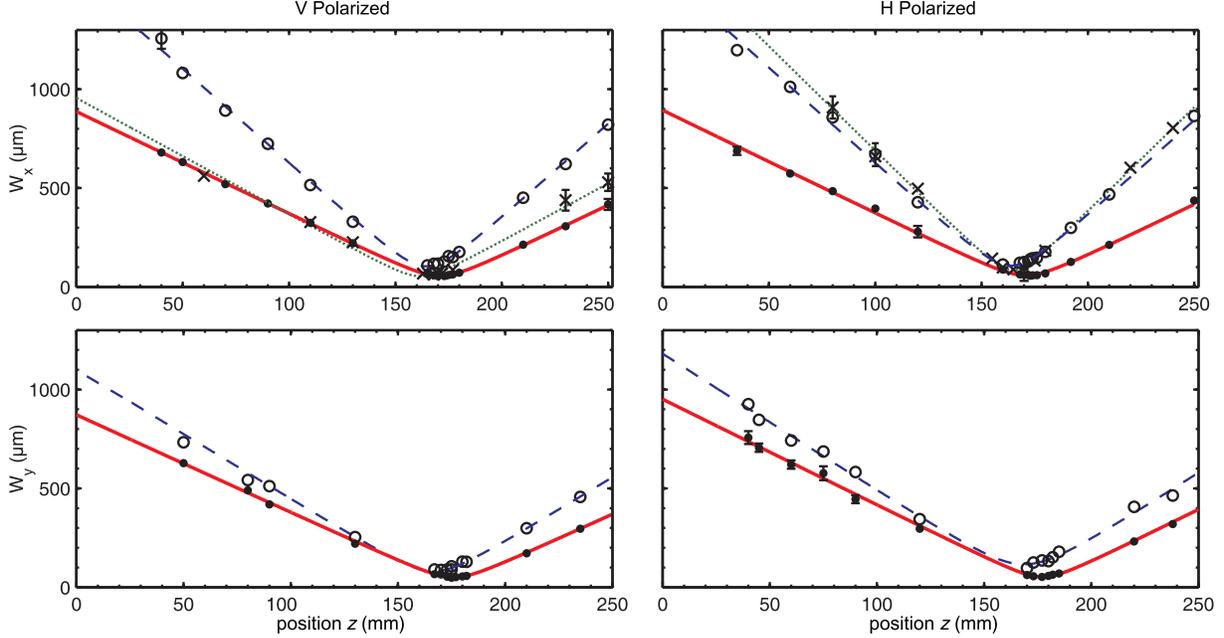}
\caption{Measurements of the beam quality factor $M^2$ for the down-converted light. Experimental results are shown as point together with lines depicting fits of the standard propagation rule for the beam width: coincidences (full circles with red solid lines), single counts (empty circles with blue dashed lines), coincidences with pump coupling optimized for efficiency instead of pump mode purity (crosses with green dotted lines).}
\label{Fig:m2}
\end{figure}

\begin{table}
\vspace*{5mm}
\centering
\begin{tabular}{lcccc}
\hline
\multicolumn{1}{c}{ Polarization   }    &  \multicolumn{2}{c}{ V } &  \multicolumn{2}{c}{ H }   \\ \cline{2-3} \cline{4-5}
 \multicolumn{1}{c}{  Scan direction }         &    vertical  $x$  &  horizontal $y$         &   vertical $x$      & horizontal $y$         \\  \hline
\multicolumn{5}{l}{{\em Pump in $00_P$ mode}} \\
Coincidences &    $1.1 \pm 0.1$   &  $0.98 \pm 0.05$    &   $1.15 \pm 0.1$    & $1.1 \pm 0.1$   \\
Single counts &    $3.9 \pm 0.4$   &  $2.1 \pm 0.3$    &  $4.1 \pm 0.4$   & $3.2 \pm 0.3$\\ \hline
\multicolumn{5}{l}{{\em Pump optimized for coupling efficiency}} \\
Coincidences &    $1.2 \pm 0.2$   &   --   &   $3.6 \pm 0.3$    & --  \\
Single counts &    $4.0 \pm 0.4$   &  --    &   $4.3 \pm 0.4$    & --  \\
\hline
\end{tabular}
\caption{Experimentally determined values of the beam quality factor $M^2$ for fields $H$ and $V$ for blade scans in two perpendicular directions in the transverse plane. }
\label{tab:M2s}
\end{table}

We took measurements of single count rates in the scanned arm, as well as coincidences with the conjugate herald photons in the reflected port of the polarizing beam splitter. The heralds were spectrally filtered with a 10~nm FWHM interference filter, coupled into a MMF using another 8~mm focal length AL, and delivered to a second SPCM. Signals from both SPCMs were directed to a coincidence circuit with a $6$~ns coincidence window. For every location $z$ and the orientation of the edge, we fitted to the experimental data the Gauss error function parameterized with the beam half-width $w$ specified at $1/e^2$ of the maximum. The coefficients of determination $R^2$ exceeded $0.998$ for all the fits. In Fig.~\ref{Fig:m2} we depict the dependence $w(z)$ for the $H$ and $V$ photons and the two edge orientations. The experimental points were fitted with a formula describing propagation of a partially coherent Gaussian beam characterized by the $M^2$ factor,\cite{QuadraticPropagation2} giving nearly unit values for coincidence measurements as specified in Tab.~\ref{tab:M2s}. These results enable us to estimate the contribution from transverse modes other than the fundamental one, based on the worst-case scenario assumption that it was caused by higher-order modes with lowest diffraction. This contribution does not exceed $5\%$ at the $95\%$ confidence level for the V polarized beam.
It is worth stressing that these results were obtained with no spectral filtering of the photons in the scanned arm. 

The $M^2$ values for single counts were much higher than $1$ for both horizontal and vertical scan directions. This is consistent with the results presented in Tab.~\ref{tab:waski_lista}, since for non-heralded photons contributions from all the phase matching bands located within approx.\ $\pm 2$~nm of the fundamental band are detected. The results obtained for single counts also verify that our measurement setup detected also contributions from higher spatial modes.

To verify the influence of pump beam mode on the spatial properties of the generated radiation, we changed the coupling of the pump beam into the waveguide to maximize the coupled power rather than to target the excitation of the fundamental pump mode. The resulting pump field intensity distribution was well described by a superposition of $00_P$, $01_P$ and $02_P$ modes. Fitting the knife-edge scans with a Gauss error function yielded much higher uncertainties for half-widths $w$, as shown in Fig.~\ref{Fig:m2}(b), which is perfectly understandable given more complex spatial profiles of the generated photons, due to excitation of phase matching bands involving higher order modes (cf.\ 
Tab.~\ref{tab:waski_lista}). Based on these data, the beam quality factor can be estimated as $M^2 > 3.5$ for the $H$ polarized heralded photons. However no significant increase of the $M^2$ value was observed for the $V$ polarized photons. This can be partially explained by the assymetry in location of the $01_P \leftrightarrow 02_V + 00_H$, and $01_P \leftrightarrow 00_V + 02_H$ phase matching bands with respect to the fundamental band. These results demonstrate the importance of proper pump beam coupling and suggest better spatial properties of the $V$ polarized photons in the case of the studied sample.

\section{GENERATION OF POLARIZATION ENTANGLEMENT}

\begin{figure}[t!]
\centering
\includegraphics[width= 16cm]{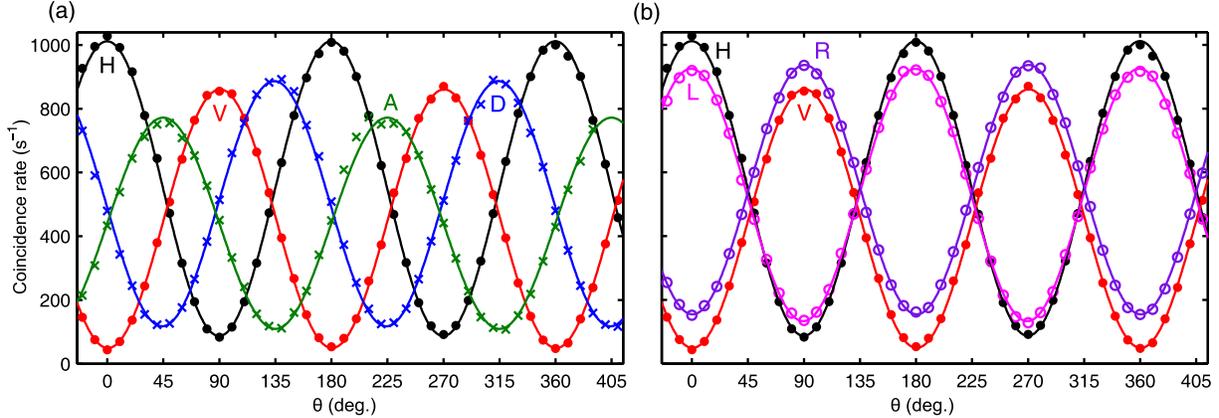}
\caption{Coincidence rates (points with error bars) with sinusoidal fits (solid lines) measured using MMFs as functions of the angle of the second polarizer. $H$, horizontal; $V$, vertical; $D$, diagonal; $A$, antidiagonal; $L$, left-circular; $R$, right-circular.}
\label{Fig:Bell}
\end{figure}

We also used our source to generate pairs of polarization-entangled photons in the Shih-Alley configuration\cite{ShihAllePRL88}. This can serve as a test of the overlap between spatial modes in which $H$ and $V$ photons have been generated.\cite{ProgOpt} The measurements were taken using the setup shown in Fig.~\ref{Fig:setup}(d). With multiphoton applications in mind, our source was pumped with femtosecond pulses which in the case of a type-II process leads to a spectral asymmetry between the generated photons. To eradicate this distinguishing information, we filtered down-converted photons using an interference filter IF4 with the bandwidth $0.7$~nm FWHM, much narrower than that needed to remove SPDC photons produced in higher-order spatial modes. In measurements described below, the waveguide was pumped with approximately 250~$\mu$W average power. After the filter, the photons were directed to a $50/50$ nonpolarizing beam splitter NPBS. Each output beam was sent through a quarter wave plate $\lambda/4$ and a half wave plate $\lambda/2$ followed by a polarizer and coupled into a fiber connected to an SPCM. The pairs of photons emerging through different output ports of the NPBS should exhibit polarization entanglement, provided that they are indistinguishable in all other degrees of freedom.

\begin{table}
\centering
\begin{tabular}{ccccc}
\hline
Reference &   \multicolumn{2}{c}{{Visibility}} \\
polarization   & SMF   & MMF  \\
\hline
H & $89\%$ & $86\%$  \\
V & $90\%$ & $90\%$  \\
D & $81\%$ & $78\%$ \\
A & $82\%$ & $77\%$  \\
R &  --    & $73\%$  \\
L &  --    & $76\%$  \\
\hline
\\
\end{tabular}
\caption{Polarization interference visibilities corrected for accidental coincidences. Values for raw data are lower by $1$-$2$ percent age points. Accidental coincidence rates were calculates using the pump repetition rate and measured single count rates.}
\label{tab:vis}
\end{table}

The coincidence rates were measured as a function of the linear polarization angle detected in one arm, with the second second polarization fixed to one of four linear cases: $0^\circ$ to select $H$ photons, $90^\circ$ to select $V$ photons, and $\pm45^\circ$ denoted as $D$ and $A$.
For measurements with multimode fibers, we also selected right- ($R$) and left- ($L$) circular polarizations in one arm and made a half-wave plate scan in the second arm for a matching orientation of the quarter wave plate.
First, as a reference measurement, we implemented spatially filtered detection using single mode fibers (SMFs) to guide the downconverted photons to the SPCMs, obtaining visibility values shown in Tab.~\ref{tab:vis}. When SMFs were replaced with MMFs with a $100~\mu$m core diameter and $0.22$ NA, only a small reduction in visibilities was observed as evidenced by the comparison in Tab.~\ref{tab:vis}. The actual coincidence count rates measured with MMFs are depicted in Fig.~\ref{Fig:Bell}, together with sinusoidal fits.
In order to confirm directly the generation of polarization entanglement we tested the Clauser-Horne-Shimony-Holt (CHSH) inequality\cite{CHSH} with standard polarizer settings.
Coincidences were recorded for $180$~s for each waveplate setting and corrected for a small drift of the pump beam power.
The obtained CHSH value of $2.319 \pm 0.006$ violates the inequality by $50$ standard deviations, clearly demonstrating the nonclassical nature of polarization correlations. We would like to stress that this result was obtained without resorting to spatial filtering of the collected photons, whereas previously reported PPKTP waveguide sources of polarization entangled photon pairs\cite{ZhongOL10} employed single mode fibers to perform spatial filtering. 

A comparison of the visibility data for SMFs and MMFs shows that the decrease of the CHSH value below the $2 \sqrt{2}$ theoretical limit is mainly caused by factors other then spatial mode mismatch.
An intriguing feature of visibility measurements for $0^\circ$ and $90^\circ$ angles is the noticeable deviation from the ideal value of $100\%$, implying occasional detection of photon pairs with identical polarizations. A possible source of these events is the contribution from higher SPDC orders, when two or more pairs are generated by the same pump pulse. Quantifying exactly this contribution was difficult as a fraction of single-photon events did not originate from the SPDC process of interest, which could be inferred from the presence of a broad background component in single-photon spectra. Under semi-quantitative assumptions, between 25\%\ and 50\%\ identical-polarization coincidences could be attributed to higher-order SPDC terms.
It is possible that some of the same-polarization photon pairs are produced by another parasitic nonlinear process that may be specific to our sample.

\section{SUMMARY}

We reported here generation of photon pairs in a multimode nonlinear waveguide using the process of spontaneous parametric down-conversion. Implementing a careful control of the spatial and spectral properties of the pump beam enabled us to produce photons with a high degree of spatial purity. This was verified by the measurement of $M^2$ beam quality factors in the heralded regime. Their values was close to the ideal value of one, in contrast to results obtained for unconditioned detection of single photons and for a misaligned pump. This clearly demonstrates that precise management of the spatial properties of down-converted photons in a multimode waveguide is feasible in practice. The basic tools are dispersive properties of the waveguide and a proper arrangement of the pump field. Because PPKTP waveguides are not easily modeled using {\em ab initio} techniques, the relevant characteristics, such as the phase matching function and the spatial shape of the pump mode, need to be measured in a separate prior experiment. Here, the technique of sum frequency generation operating on macroscopic beams comes in as a very handy utility.\cite{KarpRadzAPL09}

Development of practical tools to control fully parametric down-conversion in multimode waveguides opens up prospects to generate more complex forms of entangled states, e.g.\ exhibiting spatial entanglement. As a simple example, preparing the pump in the $01_P$ mode enables one to generate simultaneously pairs $01_H + 00_V$ and $00_H + 01_V$ with overlapping spectra\cite{KarpRadzAPL09,MoslChrisPRL09}. Nonclassical features of such a state could be tested using phase space Bell's inequalities.\cite{BanaWodkPRL99} Furthermore, this approach could be generalized to produce hyperentangled states\cite{Hyperentanglement} and combined with mode sorting techniques\cite{ModeSorting}.
Another interesting extension would to prepare polarization entanglement that would not be conditioned on paths taken by photons using bidirectional pumping\cite{Bidirectional1, Bidirectional2}, although this may be achieved most easily in a fully integrated optical circuit due to the need for an extremely precise alignment.


\acknowledgments     

We would like to acknowledge insightful discussions with C. Silberhorn and I. A. Walmsley. This work was supported by the Foundation for Polish Science TEAM project co-financed by the EU European Regional Development Fund, FP7 FET Proactive project Q-ESSENCE (Grant Agreement no.\ 248095), and the Polish Ministry of Science grant no.\ N~N202~482439.



\end{document}